\pdfoutput=1
\documentclass[aps, reprint, superscriptaddress, nofootinbib,preprintnumbers,twocolumn,showpacs,floatfix]{revtex4-1}
\usepackage{amsmath, latexsym, amssymb, graphicx, color, slashed, hyperref}
\definecolor{nicered}{rgb}{.7,.1,.1}
\definecolor{nicegreen}{rgb}{.1,.5,.1}
\definecolor{darkblue}{rgb}{0,0,.5}
\hypersetup{colorlinks, citecolor=nicegreen,linkcolor=nicered, urlcolor=darkblue}

\begin{document}

\title{Strong P invariance, neutron EDM and minimal Left-Right parity at LHC}

\author{Alessio Maiezza}
\email{amaiezza@ific.uv.es}
\affiliation{IFIC, Universitat de Val\`encia-CSIC, Apt. Correus 22085, E-46071 Val\`encia, Spain}

\author{Miha Nemev\v{s}ek}
\email{miha.nemevsek@ijs.si}
\affiliation{SISSA, Trieste, Italy}
\affiliation{INFN, Trieste, Italy}
\affiliation{Jo\v{z}ef Stefan Institute, Ljubljana, Slovenia}

\date{\today}

\begin{abstract}
In the minimal Left-Right model the choice of left-right symmetry is twofold: either generalized parity $\mathcal P$ or charge conjugation $\mathcal C$. In the minimal model with spontaneously broken strict $\mathcal P$, a large tree-level contribution to strong CP violation can be computed in terms of the spontaneous phase $\alpha$. Searches for the neutron electric dipole moments then constrain the size of $\alpha$. Following the latest update on indirect CP violation in the kaon sector, a bound on $W_R$ mass at $20 \text{ TeV}$ is set. Possible ways out of this bound require a further hypothesis, either a relaxation mechanism or explicit breaking of $\mathcal P$. To this end, the chiral loop of the neutron electric dipole moment at next-to-leading order is re-computed and provides an estimate of the weak contribution. Combining this constraint with other CP violating observables in the kaon sector allows for $M_{W_R} \gtrsim 3 \text{ TeV}$. On the other hand, $\mathcal C$-symmetry is free from such constraints, leaving the right-handed scale within the experimental reach.
\end{abstract}


\maketitle

\section{Introduction}

Left-Right(LR) symmetric theories~\cite{Pati:1974yy} offer an understanding of parity violation~\cite{Senjanovic:1975rk} and neutrino mass origin through the see-saw mechanism~\cite{seesaw}. This framework may be directly tested at the LHC via the Keung-Senjanovi\'c~\cite{Keung:1983uu} production of a heavy Majorana neutrino~\cite{Senjanovic:2010nq}. Such observation would manifest lepton number violation and Majorana nature of heavy neutrino directly at high energies with a reach of $W_R$ mass at $5-6 \text{ TeV}$~\cite{Ferrari:2000sp}.

The underlying postulate of parity restoration makes the minimal LR symmetric model (LRSM) predictive in a number of ways. It constrains the flavor structure of gauge and Higgs interactions and thus governs production at colliders, nuclear transitions such as neutrino-less double beta decay~\cite{Mohapatra:1980yp, Tello:2010am} (see also~\cite{Barry:2013xxa}), indirect constraints and early universe processes such as thermal production of warm dark matter~\cite{Bezrukov:2009th}. It ensures a direct connection between Majorana and Dirac masses, promoting LRSM to a complete theory of neutrino mass~\cite{Nemevsek:2012iq}.

The choice of LR parity however, is not unique. It can be defined either as generalized parity $\mathcal P$ or charge conjugation $\mathcal C$, see e.g.~\cite{Maiezza:2010ic}. The former may offer an insight into the strong CP problem~\cite{Mohapatra:1978fy}, while the latter can be gauged and embedded in $SO(10)$.

Indirect constraints on the LR scale have been intensely studied since the conception of LR theory. The early bound from kaon mixing~\cite{Beall:1981ze} was revisited a number of times~\cite{Mohapatra:1983ae, Zhang:2007fn} demonstrating the scale of LRSM is allowed within the reach of the LHC~\cite{Maiezza:2010ic}. A recent study~\cite{Bertolini:2014sua} updates the limit to $M_{W_R} \gtrsim 3 \text{ TeV}$ and highlights the importance of current and future constraints from $B$ physics. Regardless of how one defines parity, LR scale can be within the reach of LHC, as far as $K$ and $B$ physics is concerned.

A particularly stringent probe of P and CP violating interactions are electric dipole moments (EDM) of nucleons and atoms~\cite{edmreviews, Engel:2013lsa}. After the initial suggestion to use the neutron EDM ($n$EDM) as a probe of parity violation~\cite{Purcell:1950zz} and subsequent discovery of parity breaking in weak interactions~\cite{Lee:1956qn}, the limit from early searches~\cite{Smith:1957ht} steadily improved by around 6 orders of magnitude~\cite{Baker:2006ts}.

In the Standard Model (SM), such searches constrain the CPV $\theta$ term ($G \tilde G$) and lead to the so-called strong CP problem, a quest to explain why this parameter should be small. An attractive solution was put forth in~\cite{Peccei:1977hh} by imposing a global Peccei-Quinn (PQ) symmetry. On the other hand, since the $\theta$ term violates P (and CP), parity restoration at high scales may offer a mechanism~\cite{Beg:1978mt, Mohapatra:1978fy}, different from the usual light axion~\cite{Weinberg:1977ma}.

LR theories at TeV scales typically give a significant weak contribution to EDMs due to chirality flipping nature of gauge interactions. Short-distance effects from quark EDMs~\cite{Beall:1981zq, Ecker:1983}, the current-current operator~\cite{Ecker:1983} and the Weinberg operator~\cite{Chang:1990sfa} were studied in the past. The long distance contribution from the chiral loop was estimated in~\cite{He:1989xj}, however the result disagrees with the naive power counting~\cite{Engel:2013lsa, deVries:2012ab, Seng:2014pba}. This lead to a large limit on $M_{W_R} > 10 \text{ TeV}$ coming from the weak contribution only~\cite{Zhang:2007fn, Xu:2010}.

In this work we re-consider the issue of $n$EDM, taking into account the strong CP contribution and an updated chiral loop calculation. It is well known that a complex vev in theories with spontaneous P or CP violation introduces a tree-level contribution to $\bar \theta$~\cite{Maiezza:2010ic}. Although it vanishes in the $m_q \to 0$ limit, in the LRSM it comes out rather large for a generic spontaneous phase $\alpha$
\begin{equation} \label{eqThetaBarOne}
  \bar \theta \approx \alpha \, \left(\frac{v_2}{v_1} \right) \, \left(\frac{m_u}{m_d} \pm \frac{m_c}{m_s} \pm \frac{m_t}{m_b} \right),
\end{equation}
where $v_{1,2}$ are the usual vacuum expectation values (vevs) of the Higgs bi-doublet. The experimental EDM searches require $\alpha$ to be small, which is natural in a technical sense. As discussed in section~\ref{secwayout}, this approach differs from the one in~\cite{Maiezza:2010ic} where $\theta \neq 0$ was exploited, while exact $\mathcal{P}$ symmetry was kept in the Yukawa sector.

Following the recent results of~\cite{Bertolini:2014sua}, the indirect CP violation in the kaon sector ($\epsilon_K$) then sets the limit on the LR scale, $M_{W_R} \gtrsim 20 \text{ TeV}$. This bound can be non-trivially avoided if a relaxation PQ mechanism~\cite{Kim:1979, Shifman:1979if, Dine:1981rt} or explicit breaking in the strong sector is invoked~\cite{Maiezza:2010ic}.

%
%
\section{Minimal Left-Right Model \label{secModel}}

Left-Right symmetric theories are based on a simple extension of the SM gauge group to~\cite{Pati:1974yy}
\begin{equation} \label{eqGLR}
  SU(3)_c \otimes SU(2)_L \otimes SU(2)_R \otimes U(1)_{B-L},
\end{equation}
with LR symmetry restored at high energies. In the minimal LRSM, parity is broken spontaneously~\cite{Senjanovic:1975rk} by a pair of triplets $\Delta_L(3,1,2),\, \Delta_R(1,3,2)$ down to the SM group, followed by the final breaking with a Higgs bi-doublet
\begin{equation} \label{eqBidVev}
  \langle \Phi (2,2,0) \rangle = \text{diag} \left({v_1, e^{i \alpha} v_2 } \right).
\end{equation}
Here, $v_1^2 + v_2^2 = v^2 = (246 \text{ GeV})^2$ and $\alpha$ is the spontaneous phase.

The relevant gauge interactions for the discussion of EDMs proceed via LR gauge boson mixing
\begin{equation} \label{eqLRmix}
  \mathcal L_{LR} = \frac{g}{\sqrt 2} \, \xi \,V_R^* \, \overline u_R \slashed W d_R + \text{h.c.}.
\end{equation}
These are governed by $V_R$, the right-handed analog of the Cabibbo-Kobayashi-Maskawa (CKM) matrix $V_L$ and the mixing parameter $\xi$, the size of which depends on bi-doublet vevs
\begin{equation} \label{eqXi}
  \xi = e^{i \alpha} \sin 2 \beta \left(\frac{M_W}{M_{W_R}}\right)^2,
\end{equation}
where $\tan \beta \equiv t_\beta = v_2/v_1$. The quark Yukawa couplings can be written as~\cite{Maiezza:2010ic}
\begin{equation} \label{eqYukLR}
  \mathcal L_Y = \frac{1}{v_2} \overline Q_L \left(M \Phi + t_\beta \tilde M \tilde \Phi \right) Q_R + \text{h.c.},
\end{equation}
where $\tilde \Phi = \sigma_2 \Phi^* \sigma_2$ and after the final breaking in~\eqref{eqBidVev}, quarks become massive
\begin{equation} \label{eqMuMd}
\begin{split}
  M_u &= t_\beta^{-1} M +  t_\beta e^{-i \alpha} \tilde M,
  \\
  M_d &= e^{i \alpha} M + \tilde M.
\end{split}
\end{equation}

The discrete LR symmetry can be implemented in two ways: generalized parity or charge conjugation
\begin{equation}\label{P&C}
\mathcal{P}: \left\{ \begin{array}{l} Q_L\leftrightarrow Q_R \\[1ex]  \Phi \to \Phi^\dagger \end{array}  \right. ,
\quad
\mathcal{C}: \left\{ \begin{array}{l} Q_L \leftrightarrow (Q_R)^c \\[1ex]  \Phi \to \Phi^T \end{array}  \right. .
\end{equation}
Depending on this choice, $M$ (and $\tilde M$) is either hermitian or symmetric
\begin{equation}\label{Peffect}
  \mathcal P: M = M^\dagger, \quad \mathcal C: M = M^T.
\end{equation}

Imposing LR parity in the Yukawa sector brings about two consequences. First, the flavor structure of gauge interactions is not free.

For the case of $\mathcal C$, the right-handed mixing matrix can be written as $V_R = K_u V_L^* K_d$ where $K_{u,d}$ are arbitrary diagonal complex phases. As for $\mathcal P$, a universal $SU(3)_{L,R}$ transformation can be used to rotate e.g. $M$ to a real diagonal form and simultaneously remove two phases from $\tilde M$. This model therefore contains only two CP phases: spontaneous phase $\alpha$ and another ``hard'' phase in the Yukawa sector. We then have
\begin{equation} \label{eqVRP}
  V_R \simeq K_u V_L K_d,
\end{equation}
with external phases depending non-trivially on $\alpha$ and $\beta$~\cite{Zhang:2007fn, Maiezza:2010ic}.

The second consequence is that in the case of $\mathcal P$, $\text{arg det }M_{u,d}$ and hence $\overline \theta$ becomes calculable.

%
%
\section{Parity and the strong CP problem \label{secStrongCP}}

The standard solution to the strong CP problem is the introduction of a global PQ symmetry~\cite{Peccei:1977hh}, which provides the axion~\cite{Weinberg:1977ma} upon spontaneous breaking. The original mechanism is not phenomenologically viable, however ``invisible'' models are still allowed~\cite{Kim:1979, Shifman:1979if, Dine:1981rt}. In the SM with PQ symmetry, the axion potential relaxes at a minimum well below the experimental limit, therefore this may be seen as a dynamical explanation of small $\bar \theta$, which may also play the role of a dark matter candidate~\cite{Preskill:1982cy}.

An alternative approach to the strong CP problem is to impose P or CP symmetry, which sets the $G \tilde G$ term to zero. As long as  contributions from quark mass matrices and other CPV interactions stay below the experimentally allowed value, this approach may be considered as a solution to the strong CP problem. Such line of thought was initiated in~\cite{Beg:1978mt, Mohapatra:1978fy}, with a natural place in the context of LR symmetry~\cite{Mohapatra:1978fy}, where $\mathcal P$ acts as LR parity (for  recent work, see~\cite{Kuchimanchi}).

\subsection{$\bar{\theta}$ at tree level}
At an energy scale where $\mathcal P$ is a good symmetry, the parity violating $G \tilde G$ term is absent
\begin{equation}
  \mathcal P: \theta = 0.
\end{equation}
Such imposition is consistent with~\eqref{Peffect}, since a chiral transformation relates the strong and the weak sector~\cite{Fujikawa:1979ay}. As long as quark Yukawa couplings are hermitian, $\theta$ and $\text{arg det}(M \tilde M)$ should be small. After spontaneous breaking, all the effects parity violation are calculable and $\bar \theta$ can be computed from the determinant of quark mass matrices
\begin{equation}
  \bar \theta = \text{arg det }M_u M_d.
\end{equation}
In the LRSM with $\mathcal C$, $\theta$ remains a free parameter and $\bar \theta$ is not computable.

Conversely, for the case of $\mathcal P$, $\bar \theta$ can be approximated for a given $\alpha$ and $\beta$. Starting from~\eqref{eqMuMd}, we have
\begin{equation} \label{eqMu2}
  M_u = \left(t_\beta^{-1}-t_\beta \right) M +  t_\beta e^{-i \alpha} M_d.
\end{equation}
Neglecting the $M_d$ term provides an estimate valid up to $\mathcal O(m_b/m_t)$. In this approximation, $M$ can be rotated
\begin{equation} \label{eqMApprx}
  M = \frac{t_{2\beta}}{2} m_u s_u,
\end{equation}
such that $m_u$ is a real and diagonal matrix with arbitrary signs $s_u$~\footnote{This approximation is reliable also for the first generation provided $v_2/v_1\lesssim 0.2$. The validity of this approximation is confirmed also by matching the fit of~\eqref{eqMuMd} to~\eqref{eqMApprx}, while reproducing known CPV constraint in the literature.}. To this order, $m_u$ does not contribute to $\bar \theta$, apart from the off-set by $\pi$ due to $s_u$. For non-zero quark masses
\begin{equation} \label{eqThetaBar}
  \overline \theta = \arg \det M_d = \arg \det V_L^{\,} m_d V_R^\dagger = \text{arg det } V_R.
\end{equation}

From~\eqref{eqMuMd} we have an equation for $V_R$
\begin{equation} \label{eqVRApprx}
   V_L m_d V_R^\dagger s_u - s_u V_R^{\,} m_d V_L^\dagger = i \sin \alpha \, t_{2 \beta} \, m_u s_u,
\end{equation}
from which it is clear that $\bar \theta \propto \alpha \, (v_2/v_1)$ and the proportionality factor can be obtained from~\eqref{eqVRApprx}. In the regime $t_\beta \ll m_b/m_t$, a similar equation for $V_R$ and an analytical solution was first derived in~\cite{Zhang:2007fn}, while a general solution in the complete parameter space was recently found in~\cite{Senjanovic:2014pva}. Setting $V_L=1$ one easily recovers~\eqref{eqThetaBarOne}. Turning on the CKM mixing angles ($s_{ij}\equiv\sin\theta_{ij},\,c_{ij}\equiv\cos\theta_{ij}$), the ansatz in~\eqref{eqVRP} gives
\begin{widetext}
\begin{equation} \label{eqThetaBarEst}
  \bar \theta \simeq \sin \alpha \, t_{2\beta} \frac{
  \left(m_d m_s m_c + m_s m_b m_t  s_{12}^2 + m_d m_b m_t  c_{12}^2 \right) s_{23}^2 +
  \left(m_d m_s m_t  + m_s m_b m_c s_{12}^2 + m_d m_b m_c c_{12}^2 \right) c_{23}^2
  }{2 \, m_d m_s m_b},
\end{equation}
\end{widetext}
where sub-leading terms were omitted. Free signs in~\eqref{eqVRP} allow for a set of discrete solutions; all of them are sizeable and the smallest $\bar \theta$ is a factor of 8 below Eq.~\eqref{eqThetaBarEst} where all signs are taken positive. A numerical solution to~\eqref{eqVRApprx}, obtained by directly solving for Euler angles and external phases of $V_R$ agrees with this estimate.

A general numerical fit of the mass spectrum in~\eqref{eqMuMd} (see Appendix of~\cite{Maiezza:2010ic} for details on the fitting procedure) including the constraint $\overline \theta < \overline \theta_{\text{exp}}$ confirms the approximation in~\eqref{eqThetaBarEst}, thus $\sin \alpha \, t_{2 \beta} \to 0$ is the only way to have a small $\bar \theta$. The addition of $\overline \theta$ constraint significantly worsens the fit unless $\sin \alpha \, t_{2 \beta} \lesssim 2 m_b/m_t \, \overline \theta_{\text{exp}}$, in agreement with~\eqref{eqThetaBarEst} and disfavoring other potential minima.

%
%
\section{$n$EDM from chiral loops \label{secnEDMChL}}
The chiral loop enhancement of $n$EDM from $\overline \theta$ is known for some thirty years~\cite{Crewther:1979}. This estimate was refined~\cite{Pich:1991fq} and revisited in the context of heavy baryon effective theory~\cite{Borasoy:2000pq}. More recently, the relativistic approach with IRreg~\cite{Ottnad:2009jw} was used together with lattice estimates of the tree level contribution.

For the $\bar \theta$ contribution the leading order (LO) analysis in chiral perturbation theory~\cite{Crewther:1979, Pich:1991fq} suffices, while for the LR operator next-to-leading order (NLO) should be taken into account~\cite{He:1989xj}. Short distance contributions due to quark dipole~\cite{Beall:1981zq, Ecker:1983} and Weinberg operator~\cite{Chang:1990sfa}  are sub-leading and so is the heavy Higgs one~\cite{Chang:1992bg}.

{\em Chiral loops.} We carry out a model independent analysis employing relativistic baryon chiral perturbation~\cite{Gasser:1983yg}, together with extended-on-mass-shell (EOMS)~\cite{Fuchs:2003qc} prescription to ensure correct power counting. Following the standard notation~\cite{Scherer:2012xha}, the relevant terms in the chiral Lagrangian are
\begin{align}
  \mathcal L^{(1)} &= \overline N \left( i \slashed D - m_N + \frac{g_A}{2} \slashed u \gamma_5 \right) N,
  \\
  \mathcal L^{(2)} &= - \frac{e}{4 m_N} \left( \kappa_p \, \overline p \sigma^{\mu \nu} p
  + \kappa_n \, \overline n \sigma^{\mu \nu} n \right) F_{\mu \nu},
\end{align}
with $m_N = 938 \text{ MeV}, f_\pi = 92.4 \text{ MeV}, g_A = 1.27, \kappa_p = 1.8$ and $\kappa_n = -1.9$. The CPV pion-nucleon couplings, induced by $\bar \theta$ and the direct LR contribution are defined as
\begin{equation}
  \mathcal L_\text{CPV} = \sqrt 2 \, \overline g_+ \left(\overline n \pi^- p + \overline p \pi^+ n \right) + \overline g_n \overline n \pi^0 n.
\end{equation}

\begin{figure}
  \includegraphics[width=4cm]{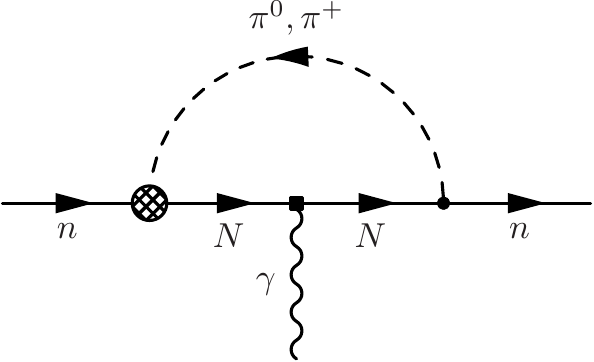}
  \raisebox{.735cm}{\includegraphics[width=4.5cm]{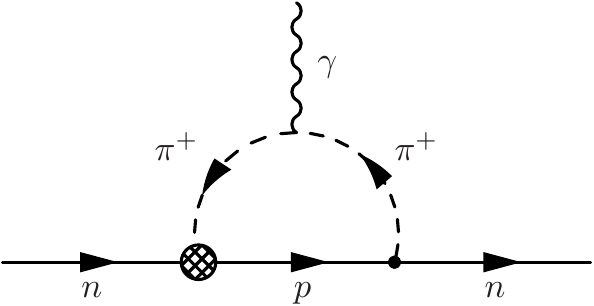}}
  \includegraphics[width=5cm]{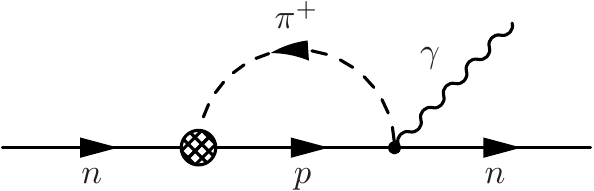}
  \caption{Loop diagrams contributing to the nucleon EDM. Hatched vertices violate CP, black square is the charge (magnetic moment) coupling of the photon to nucleons at LO(NLO) and $N=(n,p)$. The complete topology includes also the exchange of CP violating and conserving vertices.}
  \label{figLoops}
\end{figure}

One loop diagrams contributing to $n$EDM are shown in Fig.~\ref{figLoops}. The third class of diagrams cancels out while the first two give a finite contribution at LO and the topology of the first diagram gives an infinite contribution at NLO. After regularization and EOMS subtraction, the final result for the $n$EDMs is
\begin{equation} \label{eqdnChLoop}
  d_n \!= \! \frac{e}{(4 \pi)^2} \frac{g_A}{f_\pi} \left[
  \overline g_+ f(x) + \left( 2 \overline g_+ \kappa_p - \overline g_n \kappa_n \right) g(x) \right],
\end{equation}
with loop functions up to $\mathcal O(x)$, where $x = m_\pi^2/m_N^2$:
\begin{align} \label{eqf}
  f(x) & = 2 \log x - \pi \sqrt x + x,
  \\ \label{eqg}
  g(x) & = - \frac{3}{4} x \left(1+\log x \right) + \frac{3}{4} \log \left(m_N^2/\mu^2 \right).
\end{align}
The energy scale $\mu$ disappears once the (scale dependent) counter-terms are added, leading to a finite and scale independent result. Equivalently, $\log \left(m_N^2/\mu^2 \right)$ in~\eqref{eqdnChLoop} can be neglected, since we are interested in $\mu \sim 1 \text{GeV}$.

{\em Power counting.}
The expected power of a small quantity for a given diagram is~\cite{Weinberg:1991um}
\begin{equation}
  D = d - N_N - 2 N_\pi + 2 k V_\pi^{(k)} + k V_{\pi N}^{(k)},
\end{equation}
where $d$ is the dimension from loop integration, $N_N(N_\pi)$ is the number of nucleon(pion) propagators and $V^{(k)}$ is the number of vertices from the Lagrangian at a given order $k$. Diagrams on Fig.~\ref{figLoops} at LO(NLO) count as $D = 2(3)$ and since the Lagrangian term already contains one power of a small parameter, the photon momentum, the expected analytical contribution to the loop function is $D=1(2)$. Therefore, the analytic terms in Eqs.~\eqref{eqf} and~\eqref{eqg} start at $\sqrt x (x)$, as they should.

As noticed in~\cite{Engel:2013lsa}, the chiral loop calculation in early works on LR models~\cite{He:1989xj} does not obey the correct power counting and over-estimates the impact of the LR operator~\cite{He:1989xj, Xu:2010}. After an appropriate EOMS subtraction, or alternatively a computation in the heavy baryon effective theory at NNLO~\cite{Seng:2014pba}, this estimate decreases by an order of magnitude.

{\em Pion-nucleon couplings.} We proceed to estimate the CPV pion-nucleon couplings, induced by $\bar \theta$ and the LR current-current operator.

Adopting~\cite{Pich:1991fq}, the estimate of $\bar \theta$ induced couplings is
\begin{equation} \label{eqgpTheta}
  \overline g_+ = -\overline g_n = \frac{m_\pi^2}{2 f_\pi} \left( \frac{m_\Xi - m_\Sigma}{m_K^2 - m_\pi^2} \right) \, \bar \theta
  \simeq 0.05 \, \bar \theta.
\end{equation}
The chiral loop due to~\eqref{eqgpTheta} dominates the $n$EDM signal, with a lattice estimate of counter terms at $\sim 30 \%$~\cite{Ottnad:2009jw}.

Exchange of $W$ via LR gauge boson mixing in~\eqref{eqLRmix} generates LR (and RL) current-current operators
\begin{equation}
   Q_1 = \left(\overline u d \right)_{V-A} \left(\overline d u \right)_{V+A}, \,\,
   Q_2 = \left(\overline u_L u_R \right) \left(\overline d_R d_L \right),
\end{equation}
with corresponding Wilson coefficients obtained from~\eqref{eqLRmix}
\begin{equation}
  c_1(M_W) = 2 \sqrt 2 \, G_F V_{L ud}^{\,} V_{R ud}^* \, \xi, \,\, c_2(M_W) = 0.
\end{equation}
QCD running mixes the two operators, suppresses $Q_1$ and enhances $Q_2$. The anomalous matrix at NLO was obtained in~\cite{Buras:2000if} and running to 1 GeV gives
\begin{equation}
  c_1 (1 \text{ GeV}) = 0.74 \, c_1, \,\, c_2 (1 \text{ GeV}) =  -1.71 \, c_1.
\end{equation}
After a Fierz transformation and dropping the color generator terms since we are only interested in pion couplings, the CPV effective Hamiltonian can be written as
\begin{equation} \label{eqHLR}
  \mathcal H_{LR} \simeq 3 \, G_F \, c_{LR} \left[ \left(\overline u \gamma_5  u \right)
  \left(\overline d d \right) - \left(\overline u u \right) \left(\overline d \gamma_5 d \right) \right],
\end{equation}
with a dimensionless short-distance coefficient
\begin{equation}
  c_{LR} = \text{Im} \left( V_{L ud}^{\,} \, V_{R ud}^* \, \xi \right).
\end{equation}

The leading contribution to $\overline g_+$ comes from the pion vacuum expectation value $\langle \pi^0 \rangle$. From~\eqref{eqHLR}, a linear term is created in the potential of the chiral Lagrangian which induces $\langle \pi^0 \rangle$. In the vacuum saturation approximation
\begin{align}
  \langle \pi^0 \rangle = 3 \, G_F \, c_{LR} \left( \frac{f_\pi^3 m_\pi^2}{2 m_u m_d} \right).
\end{align}
The meson fields are expanded around the vev~\cite{Pich:1991fq, An:2009zh},
leading to:
\begin{align} \label{eqgbp}
  \begin{split}
  \overline g_+ &= 2 \frac{m_\pi^2}{f_\pi^2} (b_d+b_f) \left(\frac{m_d - m_u}{m_u + m_d} \right) \langle \pi^0 \rangle
  \\
  &\simeq -10^{-7} \, c_{LR},
  \end{split}
  \\
  \begin{split}
  \overline g_n &= -4 \frac{m_\pi^2}{f_\pi^2} \left(b_0 + (b_d+b_f) \frac{m_d}{m_u + m_d} \right) \langle \pi^0 \rangle
  \\
  & \simeq 2.7 \times10^{-5} \, c_{LR},
  \end{split}
\end{align}
with $b_0 = -0.517 \text{ GeV}^{-1}$, $b_d = 0.066 \text{ GeV}^{-1}$ and $b_f = -0.213 \text{ GeV}^{-1}$.

Matching~\eqref{eqHLR} to the chiral Lagrangian provides a direct source for $\overline g_n$~\cite{He:1989xj}
\begin{equation} \label{eqgbn}
\begin{split}
  \overline g_n &=
  3 \, G_F \, c_{LR} \left(\frac{f_\pi m_\pi^2}{2 m_u m_d}\right) \langle m_u \overline u u + m_d \overline d d \rangle_n
  \\
  & \simeq 2.5 \times10^{-5} \, c_{LR},
\end{split}
\end{equation}
while direct matching to $\overline g_+$ is negligible. Here, the nucleon sigma term $\sigma_{\pi N} = (m_u + m_d)/2 \langle p | \overline u u + \overline d d | p \rangle = (45 \pm 8) \text{ MeV}$~\cite{Gasser:1990ce} and mass difference $\langle p | \overline u u - \overline d d | p \rangle \simeq 0.54$~\cite{Jin:1994jz} were used to estimate the neutron matrix element.

Comparing~\eqref{eqgbp} to~\eqref{eqgbn}, we have $\overline g_+ < \overline g_n$, hence the LO contribution to $n$EDM is suppressed. Nevertheless, the chiral log compensates for this suppression and LO contributes to $d_n$ at about $60 \%$ level.

In summary, the LR current-current contribution to $d_n$ is suppressed with respect to the $\bar \theta$ one by: two orders of magnitude due to chiral matching, further two orders due to the coefficient $\text{Im}(V_{L ud}^{\,} V_{R ud}^*)$ (from the solution of $V_R$) and finally by an additional scale suppression $\xi \lesssim 10^{-3}$. $\bar \theta$ dominates the $n$EDM rate by roughly seven orders of magnitude.

\subsection{Spontaneous phase and $n$EDM}\label{spontphase}

Although $V_R$ and $\bar \theta$ are non-trivial functions of quark masses, their behaviour in the limit when either $\alpha$ or $t_\beta$ go to zero is easily understood. As shown in~\eqref{eqThetaBarEst}, $\bar \theta$ vanishes smoothly. Furthermore, from Eq.~\eqref{eqMuMd} it is clear that quark mass matrices become hermitian, such that $V_R = V_L$ with external phases $K_{u,d}$ zero or $\pi$, as evident from~\eqref{eqVRApprx}. The imaginary part of the gauge boson mixing $\xi$ disappears and so does $c_{LR}$.

Since both $\bar \theta$ and $c_{LR}$ go to zero in the same manner and $\bar \theta$ always prevails, one cannot fine-tune the two contributions. The experimental search for $n$EDM~\cite{Baker:2006ts} $d_n^{\text{exp}} < 2.9 \times 10^{-26} \text{ e cm}$ then
sets a limit on the spontaneous phase. Together with Eqs.~\eqref{eqdnChLoop} and \eqref{eqgpTheta}, one has $\bar \theta \lesssim 10^{-11}$. A somewhat relaxed limit $ \bar \theta < 1.5 \times 10^{-10}$ was obtained in a recent update~\cite{Borasoy:2000pq, Ottnad:2009jw}. Adopting the latter value, the estimate of $\bar \theta$ in Eq.~\eqref{eqThetaBarEst} translates into $\sin \alpha \, t_{2 \beta} \lesssim 10^{-11},$ such that
\begin{equation} \label{eqPnEDMVR}
  \mathcal P \, \& \, n \text{EDM} : V_R = V_L,
\end{equation}
up to arbitrary signs and to a precision of about $10^{-10}$. Note that having a small $\alpha$ is ensured once CP is imposed on the potential~\cite{Basecq:1985sx}.

\subsection{Quantum stability} The presence of low scale LR interactions might re-generate $\bar \theta$ through quantum corrections. These can be estimated using the formalism of~\cite{Ellis:1978hq}, where the SM contribution from CKM was studied.

Flavor-changing Higgs contributions are suppressed due to present constraint requiring a large mass~\cite{Maiezza:2010ic, Blanke:2011ry, Bertolini:2014sua} and multiple CKM insertions. The main effect is then due to chirality-flipping diagrams either via LR mixing or two loop $W_L-W_R$ exchange.

In the limit of~\eqref{eqPnEDMVR}, the one loop contribution due to external phases $K_{u,d}$ is
\begin{equation}
  \delta \bar \theta_{1\text{loop}} = \left( \frac{\alpha_2}{\pi} \right) \text{ Im} \left( \xi \, V_{R tb}^* V_{L tb}^{\,} \right) \left(\frac{m_t}{m_b}\right)  \lesssim 10^{-14}.
\end{equation}
At two loops, external phase contribution is further suppressed, so essentially the limit~\eqref{eqPnEDMVR} applies and the flavor structure is the same as the SM one with contributions only due to the CKM phase, as in~\cite{Ellis:1978hq}. Non-hermiticity of mass corrections then requires at least three mass insertions and the maximal amount may be estimated as
\begin{equation}
  \delta \bar \theta_{2\text{loop}} < \left( \frac{\alpha_2}{\pi} \right)^2 s_{12} s_{13} s_{23} s_\delta
  \left( \frac{m_t m_b}{M_{W_R}^2} \right) \sim 10^{-13}.
\end{equation}

In the leptonic sector, CP phases of order one alleviate the CKM suppression, however a small neutrino Dirac mass insertion is required for the chirality flip. We get
\begin{equation}
  \delta \bar \theta_{2\text{loop}} < \left( \frac{\alpha_2}{\pi} \right)^2 \left(\frac{m_t}{m_b}\right) \left( \frac{m_\tau m_D}{M_{W_R}^2} \right) \sim 10^{-14},
\end{equation}
where $m_D = \sqrt{m_\nu m_N}$ with $m_N \sim M_{W_R}$ was taken, assuming no extreme cancelation takes place in $m_D$.

Additional source of corrections to $\bar \theta$ might be due to Planck scale effects~\footnote{The explicit breaking of $\mathcal{P}$ by Planck effects is sufficient to avoid cosmologically dangerous domain walls~\cite{Rai:1992xw}.}, since quantum gravity physics might break parity. These are important for solutions based on PQ symmetry~\cite{Holman:1992us}. It turns out they are harmless for parity restoration in LRSM. As long as the effects arise from non-renormalizable operators, as in~\cite{Holman:1992us}, their impact is negligible. Parity breaking terms come first at $d=6$
\begin{equation}
  \frac{Y}{M_{Pl}^{2}} \, \overline{Q}_L \Phi Q_R \text{ tr} \left( \Delta_R^\dag \Delta_R \right) + \text{h.c.},
\end{equation}
where $Y \neq Y^{\dag}$. The suppression of non-hermitian mass correction is estimable as $v_R^2/M_{Pl}^{2}\approx 10^{-30}$~\cite{Berezhiani:1992pq}, for a right-handed scale in the TeV region.

Thus, LRSM-$\mathcal P$ in the limit of~\eqref{eqPnEDMVR} remains stable under quantum corrections.

%
%
\section{Left-Right scale: $n$EDM vs. $\varepsilon_K$ \label{secEpsK}}

The fundamental role of both direct CPV in $\varepsilon'$ and indirect in $\varepsilon_K$ for the LRSM is well known~\cite{EpsEpsKLR, epspRecent}. A recent re-evaluation of matrix elements and a discussion of relevant contributions can be found in~\cite{Bertolini:2014sua}. The authors show that one can simultaneously satisfy constraints from $K$ and $B$ mixing with the CPV ones, as long as $t_\beta > 0.02$ and $M_{W_R} > 3.2 \text{ TeV}$. In the limit of vanishing spontaneous CP violation, $\varepsilon'$ can still be satisfied, however a large bound on the LR scale emerges from $\varepsilon_K$.

As discussed above, in the minimal model the $n$EDM forces one to the limit of vanishing spontaneous CPV. For reader's convenience we recall the main issues related to $\varepsilon_K$ (for details, see~\cite{Bertolini:2014sua}) and re-compute the bound on $M_{W_R}$ in this case. To set constraints from $\varepsilon_K$, it is customary to define
\begin{equation}\label{he}
  h_\varepsilon = \frac{\text{Im}(\mathcal{H}_{LR})}{\text{Im}(\mathcal{H}_{LL})},
\end{equation}
where $\mathcal{H}_{LL}$ is the known SM effective Hamiltonian and $\mathcal{H}_{LR}$ can be written as
\begin{widetext}
\begin{equation}
\begin{split} \label{eqHepsLR}
  \mathcal H_{LR} &= G_F \sum_{i,j=c,t} \lambda_{i}^{LR} \lambda_{j}^{RL} m_i m_j
  \biggl[\frac{2 G_F}{\pi^2} \beta_W \eta_{ij}^A F_A \left(x_i, x_j, \beta_W) \right) - \frac{\eta^{LR}_{ij}}{M_H^2} \biggl( 2 \sqrt 2 +
  \\
  &\frac{G_F}{2 \pi^2} \beta_W F_C (M_{W_R}, M_H) +
  \frac{4 G_F}{\pi^2} \beta_W F_D(m_i, m_j, M_{W_R}, M_H) \biggr) \biggr] \left(\overline s_R d_L \right)\left(\overline s_L d_R \right),
\end{split}
\end{equation}
\end{widetext}
with $x_i=m_i^2/M_W^2, \beta_W = M_W^2/M_{W_R}^2$ and $\lambda_{i}^{LR(RL)}=V_{L(R) is}^{*}V_{R(L) id}^{\,}$. After imposing~\eqref{eqPnEDMVR}, the only remaining freedom for $V_R$ resides in the signs $s_i$.

\begin{figure}
  \centering
  \includegraphics[width=.7\columnwidth]{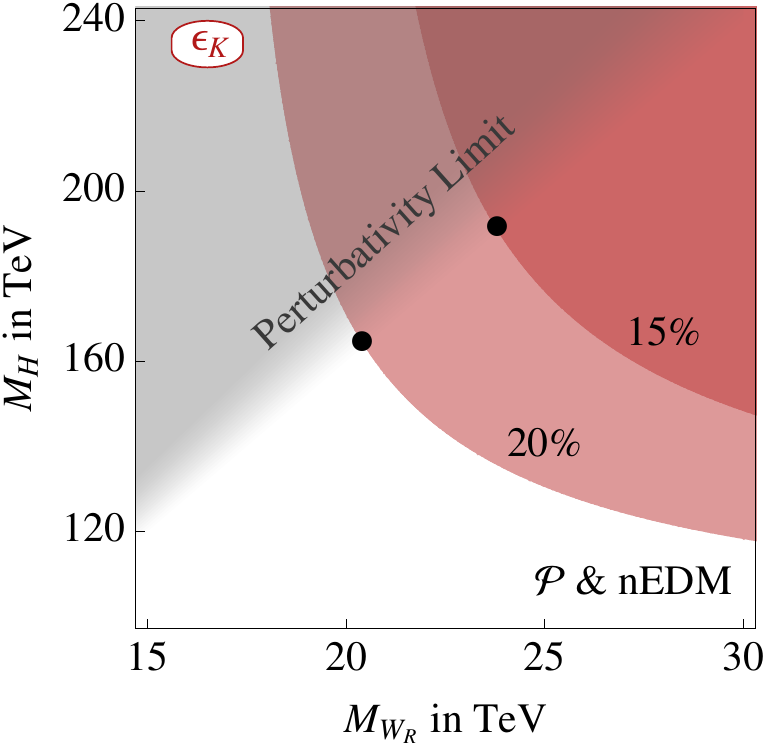}
  \caption{The bound on the LR scale in the minimal LRSM-$\mathcal P$ from $\varepsilon_K$ in the limit of vanishing spontaneous CPV. The shaded area delineates the perturbative limit, since $M_H$ and $M_{W_R}$ cannot be decoupled.}
\label{eps_vs_nEDM}
\end{figure}

As shown originally in~\cite{Basecq:1985cr}, self-energy and vertex diagrams of the heavy flavor-violating Higgs $H$ are needed for gauge invariance. Although necessary, the above contributions were neglected in the literature until~\cite{Bertolini:2014sua}, which demonstrates their importance, especially for $\epsilon_K$ and $B$ physics. The total LR contribution in Eq.~\eqref{eqHepsLR} consists of box, tree-level exchange of the heavy Higgs, self-energy and vertex diagrams. For the discussion of loop functions, QCD renormalization factors $\eta$ and the matrix elements, we remand the reader to \cite{Bertolini:2014sua}.

In the limit of~\eqref{eqPnEDMVR}, $h_\varepsilon$ depends on $M_{W_R}$ and $M_H$, with a further freedom due to the signs $s_i$. With a conservative requirement that LR contribution should not exceed $20\%$ of the SM one~\cite{Buras:2013ooa}, we find the lower limit
\begin{equation} \label{eqLRSMPbound}
  \mathcal P \, \& \, n\text{EDM} \, \& \, \epsilon_K : M_{W_R} \gtrsim 20 \text{ TeV},
\end{equation}
for the favorable choice of signs $s_c s_t=-1$ in $V_R$, as shown on Fig.~\ref{eps_vs_nEDM}.

The resulting bound applies in the minimal model with strictly imposed $\mathcal P$, where the smallness of the spontaneous phase $\alpha$ is enforced by the $n$EDM. As discussed above, this is due to $\bar \theta$ dominance, where LO dominates and the $\sim 30 \%$ uncertainty related to counter-term contributions~\cite{Ottnad:2009jw} plays a negligible role.

The huge bound in~\eqref{eqLRSMPbound} results from contradictory requirements on the spontaneous phase $\alpha$. While it is well known that a large $\alpha$ is needed to accommodate $\epsilon_K$ with TeV-scale $M_{W_R}$~\cite{Zhang:2007fn, Maiezza:2010ic, Bertolini:2014sua}, the calculable $\bar \theta$ prevents this from happening. In the following section we study two possibilities allowing to evade such tension.

%
%
\section{Ways out}\label{secwayout}

Up to this point we considered the LRSM with $\mathcal P$ broken only spontaneously and accurate to a high degree of $\sim 10^{-10}$ due to strong CP violation. In this section we discuss two separate extensions of the minimal model, avoiding the above bound.

One possibility, discussed in~\ref{subsecPQ} is to invoke the ÔÕinvisibleÕÕ axion~\cite{Kim:1979, Shifman:1979if} scenario which dynamically cancels a large strong CP phase.

Another option considered in subsection~\ref{subsecPBreak} is explicit breaking of P in the strong sector only, having $\theta \neq 0$ to cancel the effect of the spontaneous phase. This leaves us with the weak contribution only but strictly speaking leads to loss of predictability in the minimal model.

\begin{figure}
  \includegraphics[width=.75\columnwidth]{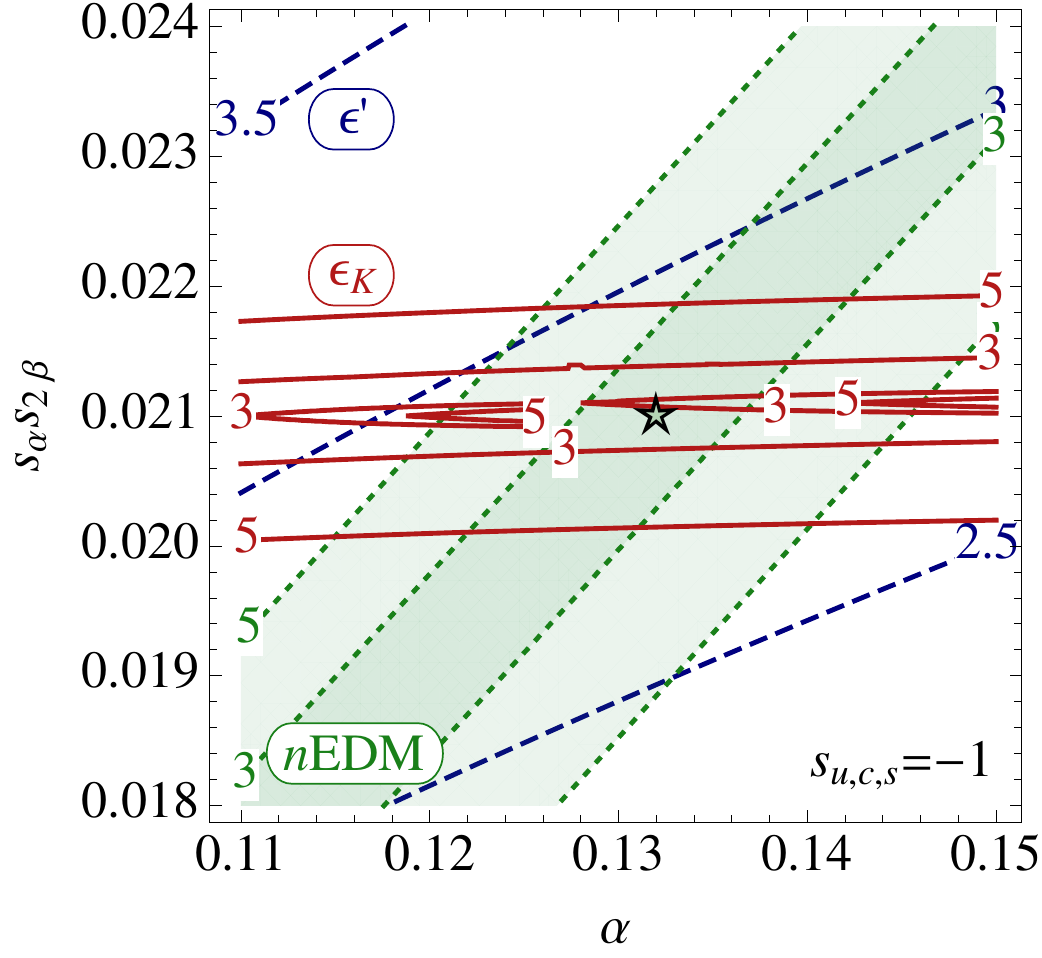}
  \caption{Combined CPV constraints in the LRSM-$\mathcal P$ extended with an ``invisible'' axion. The solution for $V_R$ obtained from~\eqref{eqVRApprx} with $s_{u,c,s}=-1$ and all others $+1$. Contours in dashed red, solid blue and dotted green show a bound on $M_{W_R}$ in TeV units coming from $\varepsilon', \varepsilon_K$ and $n$EDM via $ \bar \theta_{\text{ind}}$, respectively. The star denotes a point where all constraints are satisfied and $M_{W_R} \gtrsim 3 \text{ TeV}$.}
  \label{figEpsPnEDMEpsK}
\end{figure}

\subsection{PQ symmetry and the invisible axion. \label{subsecPQ}} Introducing a ``hidden'' PQ symmetry~\cite{Kim:1979, Shifman:1979if, Dine:1981rt} cancels away the leading strong CP term, so we are left with two weak sources of $d_n$ in LRSM. The first one is due to the chiral loop in~\eqref{eqdnChLoop}, with weak CPV vertices induced by the current-current operators in~\eqref{eqHLR}. At the same time, the presence of these CPV operators induces a linear term in the axion potential, leading to non-zero $\bar \theta$. The induced value can be estimated as~\cite{Shifman:1979if, An:2009zh}
\begin{equation} \label{eqThBarInd}
  \bar \theta_{\text{ind}} \simeq 3 \, G_F \, c_{LR} \left(\frac{f_\pi^2 m_\pi^2}{m_u m_d} \right).
\end{equation}
The contribution of $\bar \theta_{\text{ind}}$ exceeds the direct chiral loop by a factor 2-3.

At this point the issue of UV counter-terms should be pointed out. While a lattice calculation exists for $\bar \theta$~\cite{Ottnad:2009jw}, a similar result for $Q_{1,2}$ is presently unavailable. However, recent work~\cite{Seng:2014pba} suggests that, based on naive dimensional analysis, the size of the chiral loop could be $10-20\%$ of the counter-term. In such case, the direct contribution would overtake the one in~\eqref{eqThBarInd} by roughly a factor of $3-5$. Lacking a definitive conclusion, we show in Fig.~\ref{figEpsPnEDMEpsK} only the chiral loop from $\bar \theta_{\text{ind}}$ to illustrate the impact of $n$EDM. If other terms were relevant, Fig.~\ref{figEpsPnEDMEpsK} would be modified in a straightforward way. In particular, considering the counter-term estimate~\cite{Seng:2014pba} above, the green band would become narrower. This is simply because both direct and induced contributions are proportional to the same flavor structure $c_{LR} \sim \sin(\alpha-\phi_u-\phi_d)$, where $\phi_q$ are the quark phases in $K_{u,d}$. In any case, the bound $M_{W_R} \gtrsim 3 \text{ TeV}$ remains intact.

The effect of $\phi_s$ is sub-leading, it appears only when $SU(3)_f$ breaking is taken into account. However, these are quite suppressed. The main effect again is via the LO pion loop and the $\langle\eta\rangle$ in~\eqref{eqgbp}. Due to the heavier $m_s$ and $V_{us}^2 \sim 0.04$, $\phi_s$ gives at most a $10^{-3}$ correction.

The resulting bound is a combination of $\varepsilon_K, \varepsilon'$ and $\overline \theta_{\text{ind}}$. For reader's convenience, we recall that the leading contribution to $\varepsilon'$ is approximately~\cite{epspRecent, Bertolini:2014sua}
\begin{equation} \label{eqEpsp}
  \varepsilon' = 2.7 |\xi| \left( \sin(\alpha - \phi_u - \phi_d) + \sin(\alpha - \phi_u - \phi_s) \right).
\end{equation}
Clearly~\eqref{eqEpsp} has a different flavor dependence from $c_{LR}$ where the effect of $\phi_s$ is suppressed and again different from the one entering $\varepsilon_K$. This leads to a non-trivial overlap in the parameter space, shown in Fig.~\ref{figEpsPnEDMEpsK}. While the additional constraint from $n$EDM eliminates many solutions where $\varepsilon_K$ and $\varepsilon'$ agree, remarkably enough a region of parameter space can still be found, where all three constraints are in agreement, even for a low $W_R$ mass. This is further confirmed by the global fit of the mass spectrum in~\eqref{eqMuMd} with included constraints from $\varepsilon_K, \varepsilon'$ and $n$EDM.

Moreover, the numerical fit is in agreement with the low scale $M_{W_R}\simeq 3 \text{ TeV}$ in~\cite{Bertolini:2014sua}, without spoiling the phase configuration determined from $B-\overline B$ oscillation (i.e. $|\phi_b-\phi_s|\simeq|\phi_b-\phi_d|$). Although a general fit including also $B$-mixing to determine a precise bound on $M_{W_{R}}$ is beyond the scope of this work, we can conclude that PQ symmetry is an available loophole for a TeV scale $\mathcal{P}$-restoration.

\subsection{Explicit breaking of $\mathcal P$. \label{subsecPBreak}} One might entertain the idea of allowing explicit $\mathcal P$ breaking exclusively in the strong sector, while keeping the Yukawa sector intact, as in~\cite{Maiezza:2010ic}. Thus re-introduced $\theta$ then might cancel any constraint from $n$EDM~\cite{Maiezza:2010ic}. However, due to the existing search on Mercury EDM~\cite{Griffith:2009zz}, a limit from $n$EDM applies. Although conservatively the weak direct contribution does not contribute to $d_{Hg}$~\cite{Engel:2013lsa}, a limit on $\bar \theta$ from Mercury still holds. Therefore, one cannot use $\theta$ to cancel both, $\text{arg det} M_{u,d}$ and the weak contribution.

At the expense of fine-tuning and $\mathcal P$ breaking, $\theta$ can be used to cancel the phase of the quark mass determinant. Then the constraint from the pure weak chiral loop in $n$EDM remains, up to presently unknown size of the counter-terms. The flavor structure is set by the same parameter as $\bar \theta_{\text{ind}}$ but with its size suppressed by a factor $\sim 3$, so the above discussion applies and low scale $W_R$ is allowed. Note that this case is similar to the one discussed in~\cite{Zhang:2007fn, Xu:2010}, but with a different conclusion. Partly this is due to the power counting of the NLO loop function and mainly due to the fact that we find a configuration where all observables are in agreement, as shown in Fig.~\ref{figEpsPnEDMEpsK}.

This might be a conceivable way out if the amount of parity breaking were small enough not to affect other predictions due to $V_R$. Unfortunately this is not the case; the calculation of $V_R$ already demands a relatively high degree of parity invariance. Introducing a small amount of explicit $\mathcal P$ breaking in $M$ and $\tilde M$ at the level of $10^{-4}$ already affects all CPV constraints. As seen from Fig.~\ref{figEpsPnEDMEpsK} and shown in~\cite{Bertolini:2014sua}, one requires $t_\beta > 0.02$ for low scale LR. This implies $\text{arg det}\, M_u M_d \approx 0.1-1$ and in order for $\theta$ to cancel such a large contribution, parity would need to be broken by $10-100\%$. Since the non-hermiticity of quark masses and the $\theta$ term are related by the chiral transformation~\cite{Fujikawa:1979ay}, a similar amount of breaking should be allowed in the Yukawa sector as well. This departs from the minimal framework and basically allows for a free $V_R$, ruining the predictability of the model.

%
%
\section{Outlook \label{secConclude}}

The strong CP problem is addressed in the minimal LRSM with $\mathcal P$ parity. Once LR parity is imposed, nucleon EDMs are saturated by the strong contribution. For a spontaneous phase of order one, a large $\bar \theta$ in~\eqref{eqThetaBarEst} takes over the weak contribution. Setting this phase to be small simultaneously suppresses both contributions below the experimental limit, which is consistent with the minimization of the potential and remains stable under quantum corrections.

Having a small spontaneous phase in the Yukawa sector carries important consequences. Lacking additional CP phases, the bound from $\epsilon_K$ cannot be avoided and sets the LR scale to $M_{W_R} > 20 \text{ TeV}$. Such high scale is outside the reach of LHC, but may be accessible to a future generation 100 TeV collider~\cite{Rizzo:2014xma}.

A possible way out of this limit is to invoke an ``invisible'' axion mechanism or explicit $\mathcal P$ breaking. Although $n$EDM adds a non-trivial additional constraint, there still exists a portion of parameter space where low scale LR is consistent with other CPV probes, such as $\varepsilon'$ and $\varepsilon_K$. This intriguing result may change in the future, especially if CPV limits in the $B$-sector and proton (deuteron) EDM were improved~\cite{Dekens:2014jka}.

There is a consequence of having an explicit CP phase also in the leptonic sector. For the case of $\mathcal C$ parity, a direct link exists between Majorana and Dirac masses based on the symmetricity of the Dirac mass~\cite{Nemevsek:2012iq}. A similar conclusion follows for the case of $\mathcal P$. Here, the connection is less evident for an order one phase $\alpha$, but becomes clear when Dirac masses are nearly hermitian~\cite{LeptonicP}.

Regarding strong CP and other CPV constraints, the case of $\mathcal C$ parity is entirely different. There, just as in the SM, $\bar \theta$ is a free parameter and there are additional CP phases in the gauge sector. Thus, one does not need a large $t_\beta$ to accommodate $\varepsilon_K$, while $n$EDM and $\varepsilon'$ can be made small by decreasing $t_\beta$ or by proper choice of available phases in $K_{u,d}$. Either way, low scale LR parity realized as $\mathcal C$ requires no additional protection from a large $\bar \theta$ and remains to be probed at the LHC.

%
%
\section*{Acknowledgments}

\noindent We would like to thank Antonio Pich, Vladimir Tello, Yue Zhang and especially Stefano Bertolini, Fabrizio Nesti and Goran Senjanovi\'c for discussions and valuable comments on the manuscript. A.M. thanks Fabrizio Nesti for sharing his fit program. M.N. thanks Sacha Davidson, Alfredo Urbano and Felix Br\"ummer for discussions during the workshop on Axion Condensate Dark Matter at IPNL.

The work of A.M. was supported in part by the Spanish Government and ERDF funds from the EU Commission [Grants No. FPA2011-23778, No. CSD2007-00042 (Consolider Project CPAN)] and by Generalitat Valenciana under Grant No. PROMETEOII/2013/007. The work of M.N. is supported by the ERC Advanced Grant no. 267985, ÒElectroweak Symmetry Breaking, Flavour and Dark Matter: One Solution for Three MysteriesÓ (DaMeSyFla).

$\,$

\def\arxiv#1[#2]{\href{http:/arxiv.org/abs/#1}{[#2]}}


\begin{thebibliography}{99}

\bibitem{Pati:1974yy}
  J.~C.~Pati and A.~Salam,
  Phys.\ Rev.\ D {\bf 10}, 275 (1974)
  [Erratum-ibid.\ D {\bf 11}, 703 (1975)].
  R.~N.~Mohapatra and J.~C.~Pati,
  Phys.\ Rev.\ D {\bf 11}, 566 (1975).
  R.~N.~Mohapatra and J.~C.~Pati,
  Phys.\ Rev.\ D {\bf 11}, 2558 (1975).

\bibitem{Senjanovic:1975rk}
  G.~Senjanovi\'c and R.~N.~Mohapatra,
  Phys.\ Rev.\ D {\bf 12}, 1502 (1975).
  G.~Senjanovi\'c,
  Nucl.\ Phys.\ B {\bf 153}, 334 (1979).

\bibitem{seesaw}
P.~Minkowski,
Phys.\ Lett.\ B {\bf 67} (1977) 421;
R.~N.~Mohapatra, G.~Senjanovi\'{c},
Phys.\ Rev.\ Lett. {\bf 44} (1980) 912.
T.~Yanagida, {\em Workshop on unified theories and baryon number in the universe}, ed.
A. Sawada, A. Sugamoto (KEK, Tsukuba, 1979); S.~Glashow, {\em Quarks and leptons,  Carg\`ese 1979},
ed. M. L\'evy (Plenum, NY, 1980); M.~Gell-Mann \emph{et al.}, 
{\em Supergravity Stony Brook workshop, New York, 1979}, ed.\ P. Van Niewenhuizen, D. Freeman (North Holland, Amsterdam, 1980).

\bibitem{Keung:1983uu}
  W.-Y.~Keung, G.~Senjanovi\'c,
  Phys.\ Rev.\ Lett.\  {\bf 50 } (1983)  1427.

\bibitem{Senjanovic:2010nq}
For recent reviews, see
 ÊG.~Senjanovi\'c,
 Ê
 ÊInt.\ J.\ Mod.\ Phys.\ A {\bf 26} (2011) 1469
 Ê\arxiv{1012.4104}[arXiv:1012.4104 [hep-ph]];
 Ê
 ÊG.~Senjanovi\'c,
 ÊRiv.\ Nuovo Cim.\ Ê{\bf 034}, 1 (2011).
 Ê

\bibitem{Ferrari:2000sp}
  A.~Ferrari et al.
  Phys.\ Rev.\ D {\bf 62} (2000) 013001.
  S.~N.~Gninenko et al. 
  Phys.\ Atom.\ Nucl.\  {\bf 70} (2007) 441.

\bibitem{Mohapatra:1980yp}
  R.~N.~Mohapatra and G.~Senjanovi\'c,
  Phys.\ Rev.\ D {\bf 23} (1981) 165.

\bibitem{Tello:2010am}
  V.~Tello, M.~Nemev\v{s}ek, F.~Nesti, G.~Senjanovi\'c and F.~Vissani,
  Phys.\ Rev.\ Lett.\  {\bf 106} (2011) 151801
  \arxiv{1011.3522}[arXiv:1011.3522 [hep-ph]].
  M.~Nemev\v{s}ek, F.~Nesti, G.~Senjanovi\'c and V.~Tello,
  \arxiv{1112.3061}[arXiv:1112.3061 [hep-ph]].

\bibitem{Barry:2013xxa}
  J.~Barry and W.~Rodejohann,
  JHEP {\bf 1309} (2013) 153
  \arxiv{1303.6324}[arXiv:1303.6324 [hep-ph]].
  W.~-C.~Huang and J.~Lopez-Pavon,
  \arxiv{1310.0265}[arXiv:1310.0265 [hep-ph]].

\bibitem{Bezrukov:2009th}
  F.~Bezrukov, H.~Hettmansperger and M.~Lindner,
  Phys.\ Rev.\ D {\bf 81} (2010) 085032
  \arxiv{0912.4415}[arXiv:0912.4415 [hep-ph]].
  M.~Nemev\v{s}ek, G.~Senjanovi\'c and Y.~Zhang,
  JCAP {\bf 1207} (2012) 006
  \arxiv{1205.0844}[arXiv:1205.0844 [hep-ph]].

\bibitem{Nemevsek:2012iq}
  M.~Nemev\v{s}ek, G.~Senjanovi\'c and V.~Tello,
  Phys.\ Rev.\ Lett.\  {\bf 110} (2013) 15,  151802
  \arxiv{1211.2837}[arXiv:1211.2837 [hep-ph]].

\bibitem{Maiezza:2010ic}
  A.~Maiezza, M.~Nemev\v sek, F.~Nesti and G.~Senjanovi\'c,
  Phys.\ Rev.\ D {\bf 82}, 055022 (2010)
  \arxiv{1005.5160}[arXiv:1005.5160 [hep-ph]].

\bibitem{Mohapatra:1978fy}
  R.~N.~Mohapatra and G.~Senjanovi\'c,
  Phys.\ Lett.\ B {\bf 79} (1978) 283;
  K.~S.~Babu and R.~N.~Mohapatra,
  Phys.\ Rev.\ D {\bf 41} (1990) 1286.
  S.~M.~Barr, D.~Chang and G.~Senjanovi\'c,
  Phys.\ Rev.\ Lett.\  {\bf 67} (1991) 2765.

\bibitem{Beall:1981ze}
  G.~Beall, M.~Bander and A.~Soni,
  Phys.\ Rev.\ Lett.\  {\bf 48} (1982) 848.

\bibitem{Mohapatra:1983ae}
  R.~N.~Mohapatra, G.~Senjanovi\'c and M.~D.~Tran,
  Phys.\ Rev.\ D {\bf 28} (1983) 546.
  K.~Kiers, J.~Kolb, J.~Lee, A.~Soni and G.~-H.~Wu,
  Phys.\ Rev.\ D {\bf 66} (2002) 095002
  \arxiv{hep-ph/0205082}[hep-ph/0205082].

\bibitem{Zhang:2007fn}
  Y.~Zhang, H.~An, X.~Ji and R.~N.~Mohapatra,
  Phys.\ Rev.\ D {\bf 76} (2007) 091301
  \arxiv{0704.1662}[arXiv:0704.1662 [hep-ph]] and
  Nucl.\ Phys.\ B {\bf 802} (2008) 247
  \arxiv{0712.4218}[arXiv:0712.4218 [hep-ph]].

\bibitem{Bertolini:2014sua}
  S.~Bertolini, A.~Maiezza and F.~Nesti,
  \arxiv{1403.7112}[arXiv:1403.7112 [hep-ph]].

\bibitem{edmreviews}
  For recent review, see e.g.:
  J.~S.~M.~Ginges and V.~V.~Flambaum,
  Phys.\ Rept.\  {\bf 397} (2004) 63
  \arxiv{physics/0309054}[physics/0309054].
  M.~Pospelov and A.~Ritz,
  Annals Phys.\  {\bf 318} (2005) 119
  \arxiv{hep-ph/0504231}[hep-ph/0504231].

\bibitem{Engel:2013lsa}
  J.~Engel, M.~J.~Ramsey-Musolf and U.~van Kolck,
  Prog.\ Part.\ Nucl.\ Phys.\  {\bf 71} (2013) 21
  \arxiv{1303.2371}[arXiv:1303.2371 [nucl-th]].

\bibitem{Purcell:1950zz}
  E.~M.~Purcell and N.~F.~Ramsey,
  Phys.\ Rev.\  {\bf 78} (1950) 807.

\bibitem{Lee:1956qn}
  T.~D.~Lee and C.~-N.~Yang,
  Phys.\ Rev.\  {\bf 104} (1956) 254.

\bibitem{Smith:1957ht}
  J.~H.~Smith, E.~M.~Purcell and N.~F.~Ramsey,
  Phys.\ Rev.\  {\bf 108} (1957) 120.

\bibitem{Baker:2006ts}
  C.~A.~Baker {\it et al.},
  Phys.\ Rev.\ Lett.\  {\bf 97} (2006) 131801
  \arxiv{hep-ex/0602020}[hep-ex/0602020].

\bibitem{Peccei:1977hh}
  R.~D.~Peccei and H.~R.~Quinn,
  Phys.\ Rev.\ Lett.\  {\bf 38} (1977) 1440.

\bibitem{Beg:1978mt}
  H.~Georgi,
  Hadronic J.\  {\bf 1} (1978) 155;
  M.~A.~B.~B\'eg and H.~-S.~Tsao,
  Phys.\ Rev.\ Lett.\  {\bf 41} (1978) 278.

\bibitem{Weinberg:1977ma}
  S.~Weinberg,
  Phys.\ Rev.\ Lett.\  {\bf 40} (1978) 223.
  F.~Wilczek,
  Phys.\ Rev.\ Lett.\  {\bf 40} (1978) 279.

\bibitem{Beall:1981zq}
  G.~Beall and A.~Soni,
  Phys.\ Rev.\ Lett.\  {\bf 47} (1981) 552.

\bibitem{Ecker:1983}
  G.~Ecker, W.~Grimus and H.~Neufeld,
  Nucl.\ Phys.\ B {\bf 229} (1983) 421.

\bibitem{Chang:1990sfa}
  D.~Chang, C.~S.~Li and T.~C.~Yuan,
  Phys.\ Rev.\ D {\bf 42} (1990) 867.

\bibitem{He:1989xj}
  X.~-G.~He, B.~H.~J.~McKellar and S.~Pakvasa,
  Int.\ J.\ Mod.\ Phys.\ A {\bf 4} (1989) 5011
   [Erratum-ibid.\ A {\bf 6} (1991) 1063].
  X.~-G.~He and B.~McKellar,
  Phys.\ Rev.\ D {\bf 47} (1993) 4055.

\bibitem{deVries:2012ab}
  J.~de Vries, E.~Mereghetti, R.~G.~E.~Timmermans and U.~van Kolck,
  Annals Phys.\  {\bf 338} (2013) 50
  [arXiv:1212.0990 [hep-ph]].

  \bibitem{Seng:2014pba}
  C.~-Y.~Seng, J.~de Vries, E.~Mereghetti, H.~H.~Patel and M.~Ramsey-Musolf,
  \arxiv{1401.5366 }[arXiv:1401.5366 [nucl-th]].

\bibitem{Xu:2010}
  F.~Xu, H.~An and X.~Ji,
  JHEP {\bf 1003} (2010) 088
  \arxiv{0910.2265}[arXiv:0910.2265 [hep-ph]].

\bibitem{Kim:1979}
  J.~E.~Kim,
  Phys.\ Rev.\ Lett.\  {\bf 43} (1979) 103.

\bibitem{Shifman:1979if}
  M.~A.~Shifman, A.~I.~Vainshtein and V.~I.~Zakharov,
  Nucl.\ Phys.\ B {\bf 166} (1980) 493.

\bibitem{Dine:1981rt}
 M.~Dine, W.~Fischler and M.~Srednicki,
  Phys.\ Lett.\ B {\bf 104} (1981) 199.
  A.~R.~Zhitnitsky,
  Sov.\ J.\ Nucl.\ Phys.\  {\bf 31} (1980) 260
   [Yad.\ Fiz.\  {\bf 31} (1980) 497].

\bibitem{Preskill:1982cy}
  J.~Preskill, M.~B.~Wise and F.~Wilczek,
  Phys.\ Lett.\ B {\bf 120} (1983) 127,
  L.~F.~Abbott and P.~Sikivie,
  Phys.\ Lett.\ B {\bf 120} (1983) 133.
  M.~Dine and W.~Fischler,
  Phys.\ Lett.\ B {\bf 120} (1983) 137.

\bibitem{Kuchimanchi}
  R.~Kuchimanchi,
  Phys.\ Rev.\ D {\bf 82} (2010) 116008
  \arxiv{1009.5961}[arXiv:1009.5961 [hep-ph]].
  Phys.\ Rev.\ D {\bf 86} (2012) 036002
  \arxiv{1203.2772}[arXiv:1203.2772 [hep-ph]].

\bibitem{Fujikawa:1979ay}
  K.~Fujikawa,
  Phys.\ Rev.\ Lett.\  {\bf 42} (1979) 1195.

\bibitem{Senjanovic:2014pva}
  G.~Senjanovi\'c and V.~Tello,
  \arxiv{1408.3835}[arXiv:1408.3835 [hep-ph]].

\bibitem{Crewther:1979}
  R.~J.~Crewther, P.~Di Vecchia, G.~Veneziano and E.~Witten,
  Phys.\ Lett.\ B {\bf 88} (1979) 123
   [Erratum-ibid.\ B {\bf 91} (1980) 487].

\bibitem{Pich:1991fq}
  A.~Pich and E.~de Rafael,
  Nucl.\ Phys.\ B {\bf 367} (1991) 313.

\bibitem{Borasoy:2000pq}
  B.~Borasoy,
  Phys.\ Rev.\ D {\bf 61} (2000) 114017
  \arxiv{hep-ph/0004011}[hep-ph/0004011].

\bibitem{Ottnad:2009jw}
  K.~Ottnad, B.~Kubis, U.~-G.~Meissner and F.~-K.~Guo,
  Phys.\ Lett.\ B {\bf 687} (2010) 42
  \arxiv{0911.3981}[arXiv:0911.3981 [hep-ph]].
  F.~-K.~Guo and U.~-G.~Meissner,
  JHEP {\bf 1212} (2012) 097
  \arxiv{1210.5887}[arXiv:1210.5887 [hep-ph]].
  T.~Akan, F.~-K.~Guo and U.~-G.~Meissner,
  \arxiv{1406.2882}[arXiv:1406.2882 [hep-ph]].

\bibitem{Chang:1992bg}
  D.~Chang, X.~-G.~He, W.~-Y.~Keung, B.~H.~J.~McKellar and D.~Wyler,
  Phys.\ Rev.\ D {\bf 46} (1992) 3876
  \arxiv{hep-ph/9209284}[hep-ph/9209284].

\bibitem{Gasser:1983yg}
  J.~Gasser and H.~Leutwyler,
  Annals Phys.\  {\bf 158} (1984) 142.
  J.~Gasser, M.~E.~Sainio and A.~\v{S}varc,
  Nucl.\ Phys.\ B {\bf 307} (1988) 779.
  V.~Bernard, N.~Kaiser and U.~-G.~Meissner,
  Int.\ J.\ Mod.\ Phys.\ E {\bf 4} (1995) 193
  \arxiv{hep-ph/9501384}[hep-ph/9501384].

\bibitem{Fuchs:2003qc}
  T.~Fuchs, J.~Gegelia, G.~Japaridze and S.~Scherer,
  Phys.\ Rev.\ D {\bf 68} (2003) 056005
  \arxiv{hep-ph/0302117}[hep-ph/0302117].

\bibitem{Scherer:2012xha}
  S.~Scherer and M.~R.~Schindler,
  \arxiv{hep-ph/0505265}[hep-ph/0505265] and
  Lect.\ Notes Phys.\  {\bf 830} (2012) pp.1.

\bibitem{Weinberg:1991um}
  S.~Weinberg,
  Nucl.\ Phys.\ B {\bf 363} (1991) 3.

\bibitem{Buras:2000if}
  A.~J.~Buras, M.~Misiak and J.~Urban,
  Nucl.\ Phys.\ B {\bf 586} (2000) 397
  \arxiv{hep-ph/0005183}[hep-ph/0005183].

\bibitem{An:2009zh}
  H.~An, X.~Ji and F.~Xu,
  JHEP {\bf 1002} (2010) 043
  \arxiv{0908.2420}[arXiv:0908.2420 [hep-ph]],

\bibitem{Gasser:1990ce}
  J.~Gasser, H.~Leutwyler and M.~E.~Sainio,
  Phys.\ Lett.\ B {\bf 253} (1991) 252.

\bibitem{Jin:1994jz}
  X.~-m.~Jin, M.~Nielsen and J.~Pasupathy,
  Phys.\ Rev.\ D {\bf 51} (1995) 3688
  \arxiv{hep-ph/9405202}[hep-ph/9405202].

\bibitem{Basecq:1985sx}
  J.~Basecq, J.~Liu, J.~Milutinovi\'c and L.~Wolfenstein,
  Nucl.\ Phys.\ B {\bf 272} (1986) 145.
  K.~Kiers, M.~Assis and A.~A.~Petrov,
  Phys.\ Rev.\ D {\bf 71} (2005) 115015
  \arxiv{hep-ph/0503115}[hep-ph/0503115].

\bibitem{Ellis:1978hq}
  J.~R.~Ellis and M.~K.~Gaillard,
  Nucl.\ Phys.\ B {\bf 150} (1979) 141.

\bibitem{Blanke:2011ry}
  M.~Blanke, A.~J.~Buras, K.~Gemmler and T.~Heidsieck,
  JHEP {\bf 1203} (2012) 024
  \arxiv{1111.5014}[arXiv:1111.5014 [hep-ph]].

\bibitem{Holman:1992us}
  R.~Holman, S.~D.~H.~Hsu, T.~W.~Kephart, E.~W.~Kolb, R.~Watkins and L.~M.~Widrow,
  Phys.\ Lett.\ B {\bf 282} (1992) 132
  \arxiv{hep-ph/9203206}[hep-ph/9203206].
  M.~Kamionkowski and J.~March-Russell,
  Phys.\ Lett.\ B {\bf 282} (1992) 137
  \arxiv{hep-th/9202003}[hep-th/9202003].
  S.~M.~Barr and D.~Seckel,
  Phys.\ Rev.\ D {\bf 46} (1992) 539.
  S.~Ghigna, M.~Lusignoli and M.~Roncadelli,
  Phys.\ Lett.\ B {\bf 283} (1992) 278.

\bibitem{Berezhiani:1992pq}
  Z.~G.~Berezhiani, R.~N.~Mohapatra and G.~Senjanovi\'c,
  Phys.\ Rev.\ D {\bf 47} (1993) 5565
  \arxiv{hep-ph/9212318}[hep-ph/9212318].

\bibitem{EpsEpsKLR}
  G.~C.~Branco, J.~M.~Frere and J.~M.~Gerard,
  Nucl.\ Phys.\ B {\bf 221} (1983) 317,
  G.~Ecker and W.~Grimus,
  Nucl.\ Phys.\ B {\bf 258} (1985) 328, Phys.\ Lett.\ B {\bf 153} (1985) 279, Z.\ Phys.\ C {\bf 30} (1986) 293,
  G.~Barenboim, J.~Bernabeu, J.~Prades and M.~Raidal,
  Phys.\ Rev.\ D {\bf 55} (1997) 4213,
  \arxiv{hep-ph/9611347}[hep-ph/9611347],
  Y.~Rodriguez and C.~Quimbay,
  Nucl.\ Phys.\ B {\bf 637} (2002) 219,
  \arxiv{hep-ph/0203178}[hep-ph/0203178].

\bibitem{epspRecent}
  S.~Bertolini, J.~O.~Eeg, A.~Maiezza and F.~Nesti,
  Phys.\ Rev.\ D {\bf 86}, 095013 (2012),
  \arxiv{1206.0668}[arXiv:1206.0668 [hep-ph]],
  S.~Bertolini, A.~Maiezza and F.~Nesti,
  Phys.\ Rev.\ D {\bf 88}, no. 3, 034014 (2013),
  \arxiv{1305.5739}[arXiv:1305.5739 [hep-ph]].


\bibitem{Basecq:1985cr}
  J.~Basecq, L.-F.~Li and P~B.~Pal,
  Phys.\ Rev.\ D {\bf 32}, 175 (1985).

\bibitem{Buras:2013ooa}
  A.J.~Buras and J.~Girrbach,
  \arxiv{1306.3775}[arXiv:1306.3775 [hep-ph]].

\bibitem{Rai:1992xw}
  B.~Rai and G.~Senjanovi\'c,
  Phys.\ Rev.\ D {\bf 49}, 2729 (1994)
  \arxiv{hep-ph/9301240}[hep-ph/9301240].

\bibitem{Griffith:2009zz}
  W.~C.~Griffith, M.~D.~Swallows, T.~H.~Loftus, M.~V.~Romalis, B.~R.~Heckel and E.~N.~Fortson,
  Phys.\ Rev.\ Lett.\  {\bf 102} (2009) 101601.

\bibitem{Rizzo:2014xma}
  T.~G.~Rizzo,
  Phys.\ Rev.\ D {\bf 89} (2014) 095022
  \arxiv{1403.5465}[arXiv:1403.5465 [hep-ph]].

\bibitem{LeptonicP}
  M.~Nemev\v{s}ek, G.~Senjanovi\'c and V.~Tello, in preparation.

\bibitem{Dekens:2014jka}
  W.~Dekens, J.~de Vries, J.~Bsaisou, W.~Bernreuther, C.~Hanhart, U.~-G.~Mei§ner, A.~Nogga and A.~Wirzba,
  JHEP {\bf 07} (2014) 069
  \arxiv{1404.6082}[arXiv:1404.6082 [hep-ph]].

\end{thebibliography}
\end{document}